\documentclass[12pt,thmsa]{article}
\usepackage{amssymb}
\usepackage{graphicx}
\usepackage{epsf}
\usepackage{amsmath}

\begin{document}
\title{Two-Photon Effects in Lepton-AntiLepton Pair Photoproduction from
a Nucleon Target using Real Photons}
\author{Pervez Hoodbhoy \\
Department of Physics\\
Quaid-e-Azam University\\
Islamabad 45320, Pakistan. \\
{~}}
\date{March 2006}
\maketitle

\begin{abstract}
We consider the production of a lepton-antilepton pair by real photons off a
hadronic target. The interference of one and two photon exchange amplitudes
leads to a charge asymmetry term that may be calculated explicitly in the
large-$t$ limit in terms of hadronic distribution amplitudes. A rather
compact expression emerges for the leading order asymmetry at fixed angle in
the centre-of-mass of the lepton pair. The magnitude appears sizeable and is
approximately independent of the pair mass in the asymptotic limit.
\end{abstract}

Measurement of the nucleon's electric and magnetic form factors has
traditionally been based upon the well-known Rosenbluth formula\cite{Rosen}
which assumes that scattering occurs through the exchange of a single
photon. A recent extraction of $G_{E}^{p}/G_{M}^{p}$ in the $Q^{2}$ range
from 0.5 to 5.6 $GeV^{2}$ has used the polarization-transfer technique
exploited at Jefferson Laboratory\cite{JLab}. This method, which does not
use the Rosenbluth separation, revealed a large discrepancy with previously
published form factors. Subsequently, a close scrutiny was made of
two-photon exchange effects in elastic electron-proton scattering. In this
process a virtual photon knocks the incident proton into an excited state,
and a second one de-excites it back into the ground state. Both photons may
be hard, and hence they probe nucleon structure. The effects were found to
exist at a few percent level and are capable of resolving the observed
discrepancy\cite{Blunden}. Model dependence is inevitable. Chen et al. \cite%
{Chen} have also calculated the two-photon exchange contribution and related
it to the generalized parton distributions\cite{Ji} that occur in various
other hard processes as well. A clear exposition of experimental techniques
and two-photon physics may be found in a review by Wright and Jager\cite%
{Wright}.

In this paper, we shall consider two-photon effects in the
production of a lepton-antilepton pair from a hadronic target by
real photons (see Fig. 1) at high centre of mass energy $s$.

\begin{figure}
\includegraphics[height=2.5in,width=3.0in]{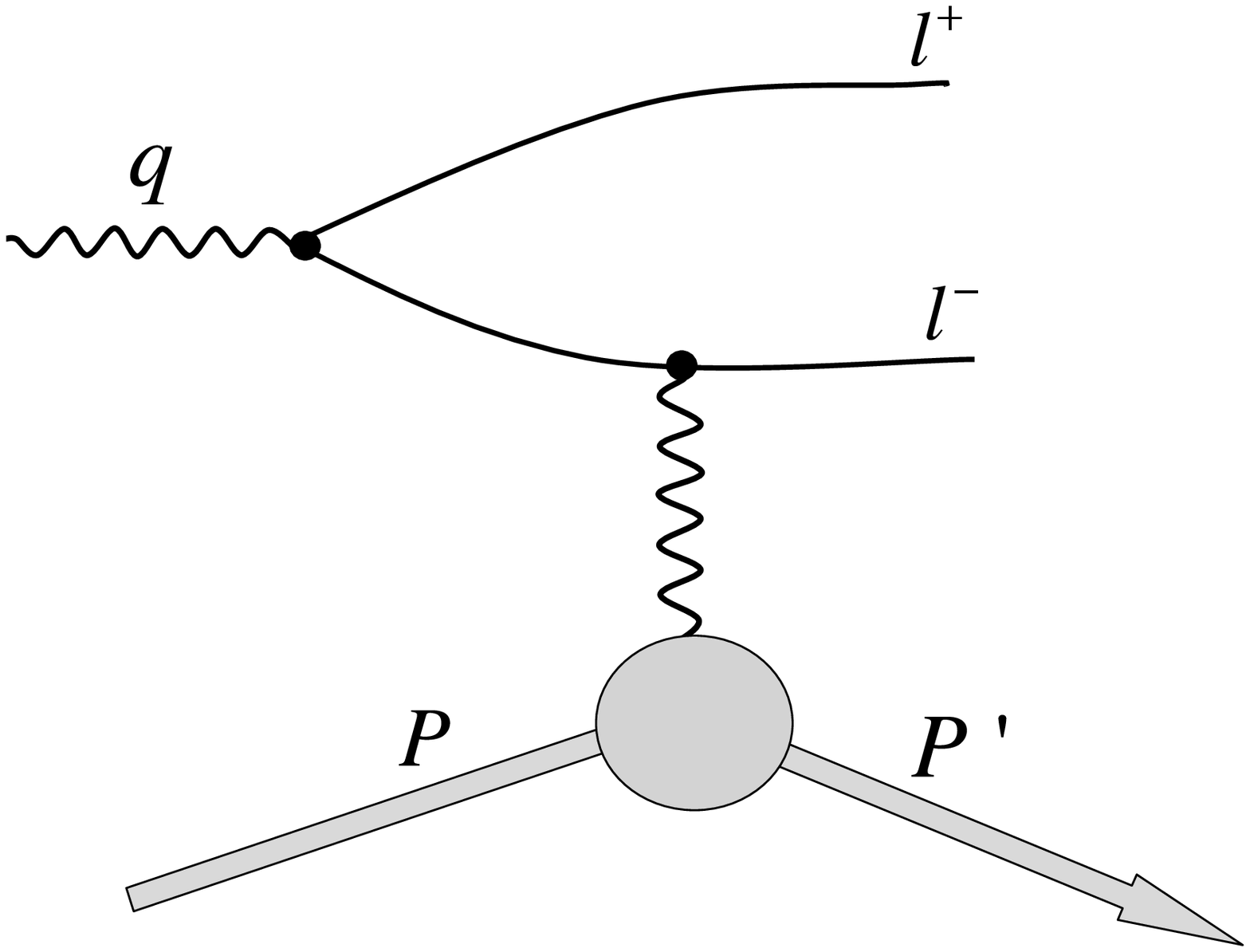}
\caption{\baselineskip = 1.0em  Exclusive photoproduction of a
lepton pair off a proton target.} \label{Fig.1}
\end{figure}

The process predominantly occurs through the exchange of one
photon,with a two-photon admixture. Both amplitudes are for the same
final state and so interfere with one another. These exchanges come
from diagrams with opposite charge conjugation \cite{Gillespie}.
This fact enables one to isolate the interference term by counting
the difference in the number of produced antileptons and leptons. In
the limit of large momentum transfer $t$, they can be explicitly
computed from the leading Fock-state in terms of lowest order
hadronic distribution amplitudes. We find a rather compact
expression for the leading order asymmetry at fixed angle in the
centre-of-mass of the lepton pair. The inputs to this calculation,
apart from the hadron distribution amplitudes,
are experimentally determined hadronic form factors at sufficiently large $t$%
. We note that Berger, Diehl, and Pire had analyzed and estimated
exclusive lepton pair photoproduction (at small $t$) as a means of
studying generalized parton distributions in the nucleon\cite{Pire}.
Exclusive electroproduction of J/psi mesons (which then decay into
lepton pairs) has been studied at HERA\cite{ZEUS} for fairly small
$Q^{2}$ values.

Before considering the more complex case of the proton, we shall first
consider a pion target. It is convenient to work in the rest frame of the
produced lepton-antilepton pair. The mass of the quarks, and of the hadronic
target, have been ignored in this preliminary calculation, i.e. $%
P^{2}=P^{\,\prime \,2}=0$ where $P^{\mu }$ and $P^{\,\prime \mu \,}$
are the momenta of the initial and final hadron. The squared
invariant mass of the produced leptons (also assumed massless) is
$M^{2}$. This may be selected at will and should be chosen far away
from a resonance. Simple expressions emerge only for
$s>>-t,M^{2}>>\Lambda _{QCD}^{2}$ where $t$ is defined, as usual, from $%
t=(P-P^{\,\prime })^{2}.$ It is negative in the physical region. The
incoming real photon ($q^{2}=0$) is taken along the $z$-axis, and
the $x-z$ scattering plane is defined by the incoming vectors $q^{\mu }$
and $P^{\mu }$
outgoing, elastically scattered, hadron is in the $x-z$ plane. The sum of
the lepton-antilepton momentum vectors is $K^{\mu }=l^{\mu }+l^{^{\prime
}\mu }$. A convenient parametrization of the scattering kinematics is then
provided by,

\begin{eqnarray}
q^{\mu } &=&(\omega ,0,0,\omega ), \\
P^{\mu } &=&(\epsilon ,\epsilon \sin \psi ,0,\epsilon \cos \psi ), \\
P^{\,\prime \,\mu } &=&(\epsilon +\omega -M,\epsilon \sin \psi ,0,\omega
+\epsilon \cos \psi ), \\
K^{\mu } &=&(M,0,0,0),
\end{eqnarray}
where the incoming hadron energy $\epsilon $, incoming photon energy
$\omega$, and angle $\psi $ are,

\begin{eqnarray}
\epsilon &=&\frac{s+t}{2M}, \\
\omega &=&\frac{M^{2}-t}{2M}, \\
\cos \psi &=&1-\frac{s}{2\omega \epsilon }.
\end{eqnarray}%

The outgoing lepton and antilepton, also taken to be massless, have spinors $%
\overline{u}(l)$ and v$(l^{^{\,\prime }}).$ The lepton and anti-lepton
momentum vectors are,

\begin{eqnarray}
l^{\,\mu } &=&\frac{M}{2}(1,\sin \theta \cos \phi ,\sin \theta \sin \phi
,\cos \theta ), \\
l^{^{\prime }\,\mu } &=&\frac{M}{2}(1,-\sin \theta \cos \phi ,-\sin \theta
\sin \phi ,-\cos \theta ).
\end{eqnarray}%
The azimuthal angle $\phi $ is measured relative to the plane formed
by $\overrightarrow{P}$ and $\overrightarrow{q}$ (which defines the
z-axis and hence $\theta $). In the centre-of-mass frame used here
$\overrightarrow{P}^{\,\prime }$ also lies in this plane. By angular
momentum conservation, the two massless leptons have opposite
helicities. Since we shall work at the amplitude level, we need
simple, covariant expressions for the matrix v$(l^{^{\,\prime
}})\overline{u}(l)$ in the helicity basis. The method developed by
Vega and Wudka\cite{Vega} is especially convenient when used in the
cm frame:

\begin{eqnarray}
\text{v}_{\downarrow }(l^{^{\,\prime }})\overline{u}_{\uparrow }(l) &=&-%
\frac{M}{4}\eta _{+}\!\!\!\!\!\!/\;\;+\frac{M}{4}\eta
_{+}\!\!\!\!\!\!/\,\;\gamma _{5}, \\
\text{v}_{\uparrow }(l^{^{\,\prime }})\overline{u}_{\downarrow }(l) &=&-%
\frac{M}{4}\eta _{-}\!\!\!\!\!\!/\;\;-\frac{M}{4}\eta
_{-}\!\!\!\!\!\!/\,\;\gamma _{5}.
\end{eqnarray}
since it will not enter the cross-sections. The auxiliary vector
$\eta _{\pm }^{\mu }$ is defined as,

\begin{equation}
\eta _{\pm }^{\mu }=(0,\cos \theta \cos \phi \pm i\sin \phi ,\cos \theta
\cos \phi \mp i\cos \phi ,-\sin \theta ).
\end{equation}%
Here $\pm $ refers to the lepton helicity. $\eta ^{\mu }$ satisfies,

\begin{eqnarray}
\eta _{-} &=&\eta _{+}^{\ast }, \\
\eta _{+}\cdot \eta _{+} &=&\eta _{-}\cdot \eta _{-}=0, \\
\eta _{\pm }\cdot l &=&\eta _{\pm }\cdot l^{^{\,\prime }}=0.
\end{eqnarray}

Consider now lepton pair production from a pion via the form factor diagram
(Fig.2). This, together with its crossed counterpart, is easily calculated
in the large $s$ limit.

$\smallskip $With the pion factor normalized such that $F_{\pi }(0)=1$, the
amplitude for producing a positive helicity lepton from a positive helicity
real photon is,

\begin{figure}
\includegraphics[height=2.5in,width=3.0in]{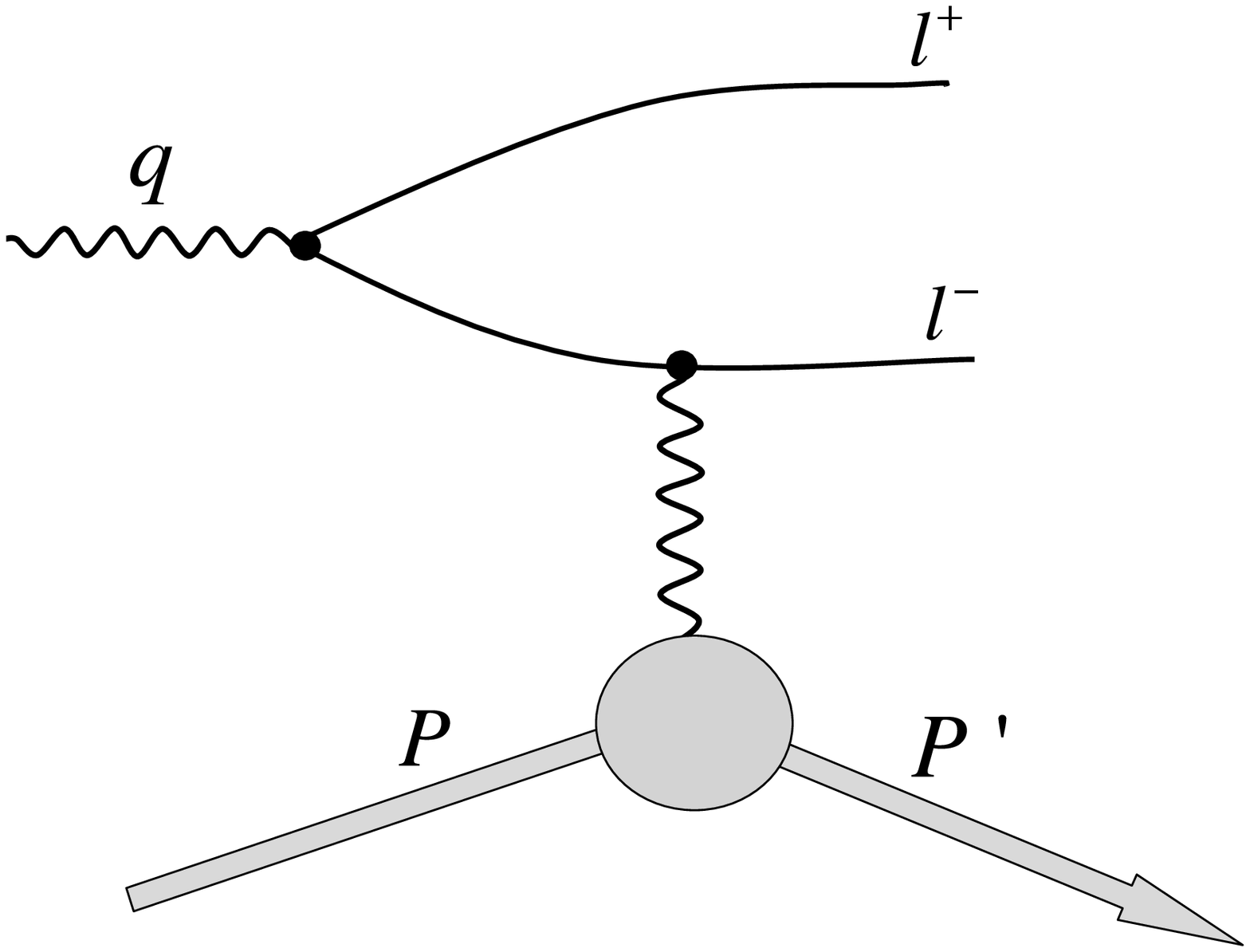}
\caption{\baselineskip = 1.0em  Form-factor contribution to lepton
pair production at lowest order.} \label{Fig.2}
\end{figure}

\begin{eqnarray}
A_{1}^{\uparrow \uparrow } &\equiv &A_{1}(\theta ,\phi ,\gamma ^{\uparrow
},l^{\uparrow },\bar{l}^{\downarrow })  \notag \\
&=&-i\sqrt{2}e^{3}\frac{s}{M^{3}\sqrt{\rho }(1+\rho )^{2}}e^{-i\theta }\left[
i\sqrt{\rho }(e^{i\theta }-1)+e^{i\phi }(e^{i\theta }+1)\right] ^{2}\csc
\theta \;F_{\pi }(t)  \notag \\
&&  \label{z1}
\end{eqnarray}%
For convenience we have defined the dimensionless momentum transfer,
\begin{equation}
\rho =-\frac{t}{M^{2}}.
\end{equation}%
By examining the $\gamma $-matrix structure, the other helicity amplitudes
are easily obtained from,

\begin{eqnarray}
A_{1}^{\downarrow \uparrow } &\equiv &A_{1}(\theta ,\phi ,\gamma
^{\downarrow },l^{\uparrow },l^{\prime \downarrow })=A_{1}(\theta +\pi
,-\phi ,\gamma ^{\uparrow },l^{\uparrow },l^{\prime \downarrow }), \\
A_{1}^{\uparrow \downarrow } &\equiv &A_{1}(\theta ,\phi ,\gamma ^{\uparrow
},l^{\downarrow },l^{\prime \uparrow })=A_{1}(\theta +\pi ,\phi ,\gamma
^{\uparrow },l^{\uparrow },l^{\prime \downarrow }), \\
A_{1}^{\downarrow \downarrow } &\equiv &A_{1}(\theta ,\phi ,\gamma
^{\downarrow },l^{\downarrow },l^{\prime \uparrow })=A_{1}(\theta ,-\phi
,\gamma ^{\uparrow },l^{\uparrow },l^{\prime \downarrow }).
\end{eqnarray}%
For momentum transfers much higher than the invariant mass of the produced
lepton pair
\begin{equation}
A_{1}^{\uparrow \uparrow }\rightarrow -i\frac{\sqrt{8}e^{3}}{(-t)^{3/2}}%
s\tan \frac{\theta }{2}F_{\pi }(t)\;\;\;\;(s>>-t>>M^{2}).
\end{equation}%
Note that the $\phi $ dependence entirely disappears in this limit. The
singular behaviour for $\theta \rightarrow 0$ comes from the lepton
propagator in Fig.2 and disappears upon including the lepton mass. However,
for purposes of comparing with the other amplitudes to be computed below,
where including the mass would make the formulae less transparent, this mass
will be kept at zero in this preliminary calculation.

The squared amplitude from Fig.2, summed over lepton polarizations, and
averaged over photon polarizations, is:
\begin{equation}
\sum_{\gamma ,s}\left| A_{1}^{\gamma s}\right| ^{2}=\frac{32s^{2}e^{6}F_{\pi
}^{2}(t)}{M^{6}\rho (1+\rho )^{4}}\sum_{n=0}^{2}a_{n}\cos n\phi .
\label{zeroth}
\end{equation}%
The coefficients $a_{n}$ are ,
\begin{eqnarray}
a_{0} &=&-(\rho ^{2}-4\rho +1)+2(\rho ^{2}+1)\csc ^{2}\theta , \\
a_{1} &=&4\sqrt{\rho }(\rho -1)\cot \theta , \\
a_{2} &=&2\rho .
\end{eqnarray}%
Under spatial inversion (i.e. $\theta \rightarrow \theta +\pi $ and $\phi
\rightarrow \phi $ ) the outgoing lepton and anti-lepton are exchanged. But $%
\frac{d\sigma }{dt}\;$, computed from Eq. \ref{zeroth} suffers no change. In
other words the charge asymmetry at leading order is zero.

A fermion line attached to three vector vertices has the opposite charge
conjugation properties relative to the same line with two vertices. We shall
use this property in an essential way. So, now consider scattering into the
same final state but through the exchange of two photons. The three photon
vertices may be connected to the $l^{-}l^{+}$ line in six different ways,
two of which have been shown in Fig.3.

\begin{figure}[t]
\includegraphics[height=2.0in,width=5.5in]{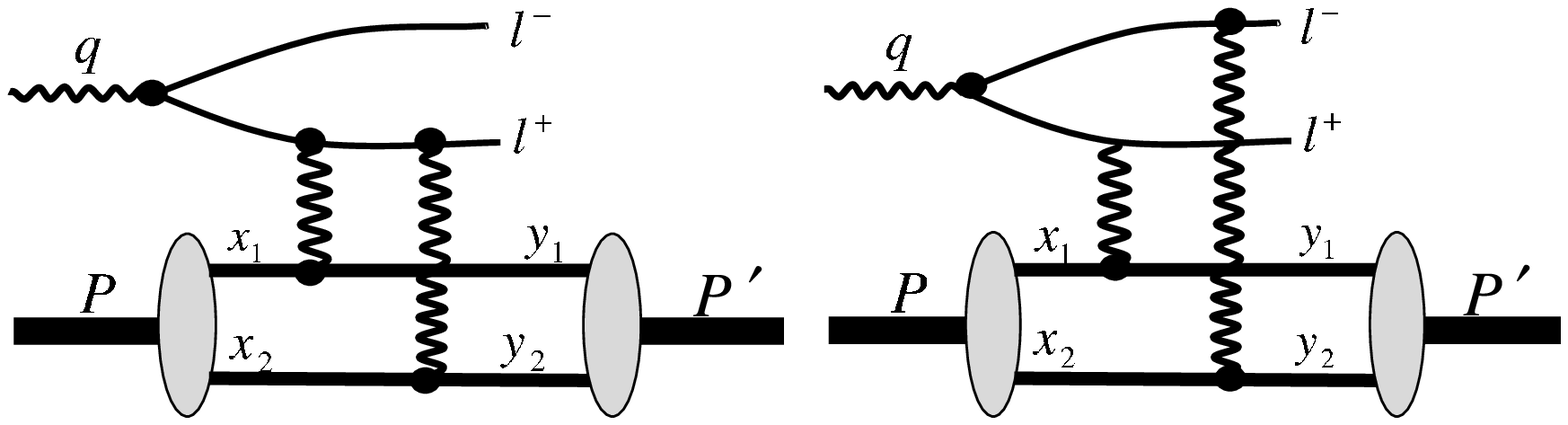}
\caption{\baselineskip = 1.0em  Typical two-photon contributions to
lepton photoproduction from a pion target.}
\label{Fig.3}
\end{figure}

In general, the scattering from any hadronic state is extremely complicated
and, in the language of Fock states, involves an infinite number of proton
wavefunction components. However, in the large-$t$ limit, all higher
components beyond the minimal one are strongly suppressed in a purely
exclusive process. The reason is straightforward: every quark and gluon
present in the initial state must be ``turned around'' and its momentum
components redirected into the final state. This simple fact allows for
calculation of hadronic form-factors\cite{Brodsky}, as well as generalized
parton distributions\cite{hoodbhoy}, etc.

In the calculation reported below we assume that large enough values of $t$
can be selected for the minimal 2-quark component of the pion to dominate
the scattering. Integrating over the transverse momentum of quarks, one may
then approximately represent the pion state by,

\begin{eqnarray}
|\pi ^{+},P\rangle &=&\int \frac{\left[ dx\right] }{\sqrt{4x_{1}x_{2}}}\Phi
(x_{1},x_{2})\frac{1}{\sqrt{3}}\left\{ u_{a\uparrow }^{\dagger }(x_{1})%
\overline{d}_{a\downarrow }^{\dagger }(x_{2})-u_{a\downarrow }^{\dagger
}(x_{1})\overline{d}_{a\uparrow }^{\dagger }(x_{2})\right\} |0\rangle ,
\notag \\
&&
\end{eqnarray}%
Where the integration measure is,
\begin{equation}
\left[ dx\right] =dx_{1}dx_{2}\delta (1-x_{1}-x_{2}).
\end{equation}%
Here $a=1,2,3$ denotes colour. Asymptotically, as is well known, $\phi _{\pi
}(x)$ approaches $f_{\pi }\sqrt{6}x(1-x).$

The calculation of diagrams in Fig.3 is straightforward. Typically one has a
denominator that, expanded out in the large $s$ limit, looks like,%
\begin{equation*}
\frac{1}{s(x_{1}-y_{1})\left[ 1+\rho +2\sqrt{\rho }\cos \phi \sin \theta
+(\rho -1)\cos \theta \right] +i\epsilon }.
\end{equation*}%
This is singular at $x_{1}=y_{1}$, and so has a principal part in addition
to an imaginary (delta function) piece. The principal parts cancel when
taking the sum of all six diagrams for the upper diagrams. The numerator is
of $O(s^{2}),$ and so the overall contribution is proportional to $s$.

After a tedious calculation, the amplitude for producing a positive helicity
lepton from a positive helicity real photon via two-photon exchange reads,
\begin{eqnarray}
A_{2}^{\uparrow \uparrow } &\equiv &A_{2}^{\gamma \pi \rightarrow \pi
l^{-}l^{+}}(\theta ,\phi )  \notag \\
&=&-\frac{2i\sqrt{2}\pi e_{u}e_{d}e^{3}s\text{ }}{M^{5}\rho ^{3/2}(1+\rho
)^{2}}e^{-i\theta }(\cos \theta +e^{-i\phi }\frac{M}{\sqrt{-t}}\sin \theta )%
\left[ i\sqrt{\rho }(e^{i\theta }-1)+e^{i\phi }(e^{i\theta }+1)\right]
^{2}\times  \notag \\
&&\;\int \left[ dx\right] \left[ dy\right] \frac{\Phi (x_{1},x_{2})\Phi
(y_{1},y_{2})\delta (y_{1}-x_{1})}{x_{1}\overline{x}_{1}\left[ x_{1}%
\overline{x}_{1}+\Lambda \right] }.
\end{eqnarray}%
In the above, $\overline{x}_{1}=1-x_{1}$ and the integrand depends upon $%
\theta $ and $\phi :$
\begin{equation}
\Lambda =\frac{e^{-i\phi }}{4\sqrt{\rho }}\sin 2\theta -\frac{1}{4}(1-\frac{%
e^{-i2\phi }}{\rho })\sin ^{2}\theta .
\end{equation}%
The other amplitudes can be expressed in terms of $A_{2}^{\uparrow \uparrow
}:$%
\begin{eqnarray}
A_{2}^{\uparrow \downarrow } &\equiv &A_{2}(\theta ,\phi ,\gamma ^{\uparrow
},l^{\downarrow },\bar{l}^{\uparrow })=-A_{2}(\theta +\pi ,\phi ,\gamma
^{\uparrow },l^{\uparrow },\bar{l}^{\downarrow }), \\
A_{2}^{\downarrow \uparrow } &\equiv &A_{2}(\theta ,\phi ,\gamma
^{\downarrow },l^{\uparrow },\bar{l}^{\downarrow })=-A_{2}(\theta +\pi
,-\phi ,\gamma ^{\uparrow },l^{\uparrow },\bar{l}^{\downarrow }), \\
A_{2}^{\downarrow \downarrow } &\equiv &A_{2}(\theta ,\phi ,\gamma
^{\downarrow },l^{\downarrow },\bar{l}^{\uparrow })=A_{2}(\theta ,-\phi
,\gamma ^{\uparrow },l^{\uparrow },\bar{l}^{\downarrow }).  \label{two-1}
\end{eqnarray}%
The negative signs in the above amplitudes will be responsible for the
charge asymmetry in the crossection, as we shall see shortly. But first, let
us remark on the apparent problem in the integral,
\begin{equation}
\int \left[ dx\right] \left[ dy\right] \frac{\Phi (x_{1},x_{2})\Phi
(y_{1},y_{2})\delta (y_{1}-x_{1})}{x_{1}\overline{x_{1}}\left[ x_{1}%
\overline{x_{1}}+\Lambda \right] }.
\end{equation}%
For $\phi =0$ the integrand is real and has a singularity inside the
integration range. However, reinstating the lepton mass removes the
singularity by introducing a term of order $m^{2}/M^{2}$. This still leaves
a very strong $\phi $ dependence, thereby distinguishing the two-photon
exchange term from that with a single photon. A proper calculation must, of
course, include lepton masses. One might wonder about other diagrams, also
of $O(e^{3}),$ such as those in Fig.4. However, these can be shown to be of $%
O(1/s)$.

\begin{figure}
\includegraphics[height=2.0in,width=3.0in]{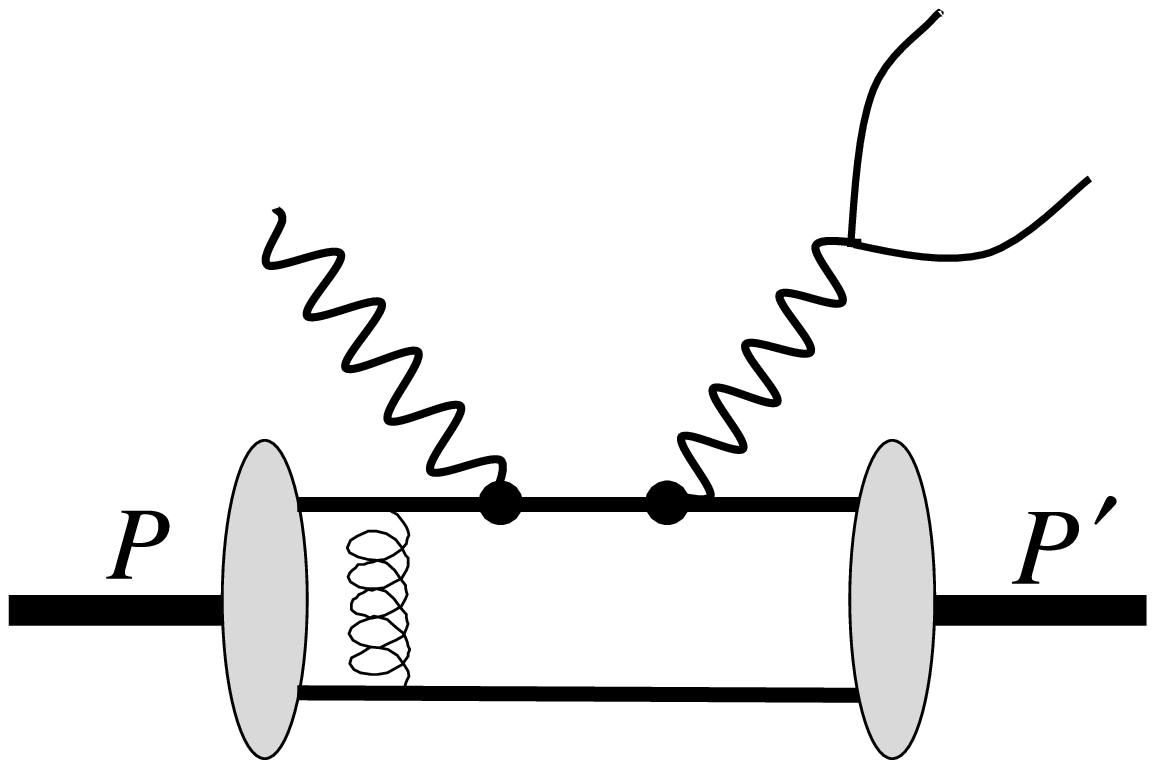}
\caption{\baselineskip = 1.0em  A typical diagram for lepton
pairproduction from a single quark in the target. The sum of all
such diagrams is suppressed in the large - s limit for real
photons.}
\label{Fig.4}
\end{figure}

Now consider the interference term in $\left| A_{1}+A_{2}\right|^{2}$,

\begin{equation}
\Theta (\theta ,\phi )=\frac{\sum_{\gamma ,s}(A_{1}^{\gamma s}A_{2}^{\ast
\gamma s}+A_{1}^{\ast \gamma s}A_{2}^{\gamma s})}{\sum_{\gamma ,s}\left|
A_{1}^{\gamma s}\right| ^{2}}.  \label{asy}
\end{equation}%
After simplication, this becomes,

\begin{eqnarray}
\Theta (\theta ,\phi ) &=&\frac{16\pi ^{2}e_{u}e_{d}\alpha _{e}}{M^{2}\rho
F_{\pi }(t)}(\cos \theta +e^{-i\phi }\frac{1}{\sqrt{\rho }}\sin \theta )
\notag \\
&&\times \int \left[ dx\right] \left[ dy\right] \frac{\Phi (x_{1},x_{2})\Phi
(y_{1},y_{2})\delta (y_{1}-x_{1})}{x\overline{x}\left( x\overline{x}+\Lambda
\right) }+cc,  \notag \\
&=&\frac{16\pi ^{2}e_{u}e_{d}\alpha _{e}}{M^{2}\rho F_{\pi }(t)}(\cos \theta
+e^{-i\phi }\frac{1}{\sqrt{\rho }}\sin \theta )\int_{0}^{1}dx\frac{\Phi
^{2}(x)}{x\overline{x}\left( x\overline{x}+\Lambda \right) }+cc.  \notag \\
&&
\end{eqnarray}%
In the above, $\overline{x}=1-x=1-x_{1}$ and $e_{u}=2/3$, $e_{d}=-1/3$. We
define the charge asymmetry as:

\begin{equation}
\Xi (\theta ,\phi )=\Theta (\theta +\pi ,\phi )-\Theta (\theta ,\phi
)=-2\Theta (\theta ,\phi ).
\end{equation}%
Obviously $\Xi (\theta ,\phi )=-\Xi (\theta +\pi ,\phi )$. This quantity is
proportional to the difference in count rates between antileptons and
leptons. As is apparent from the definition of $\Lambda $, the integral over
$x$ asymptotically goes to a constant, finite value as $\rho \rightarrow
\infty .$ A leading order calculation\cite{Brodsky},\cite{Efremov} for the
pion form factor gives $(-t)F_{\pi }=12f_{\pi }^{\;2}\pi C_{F}\alpha _{S}$\
at large $t$ or, equivalently, $\rho F_{\pi }=12(f_{\pi }^{\;2}/M^{2})\pi
C_{F}\alpha _{S}$. Note that $M$, the lepton pair invariant mass, has
disappeared from the final formula for $\Xi .$\ In Fig.5, $\Xi $ is plotted
as a function of $\rho $ for fixed angles. Note that we have assumed
collinear quarks and so the formula is valid only for $M$ greater than the
typical $k_{T}$ of quarks inside the pion.

\begin{figure}
\includegraphics[angle=90,height=3.5in,width=5.0in]{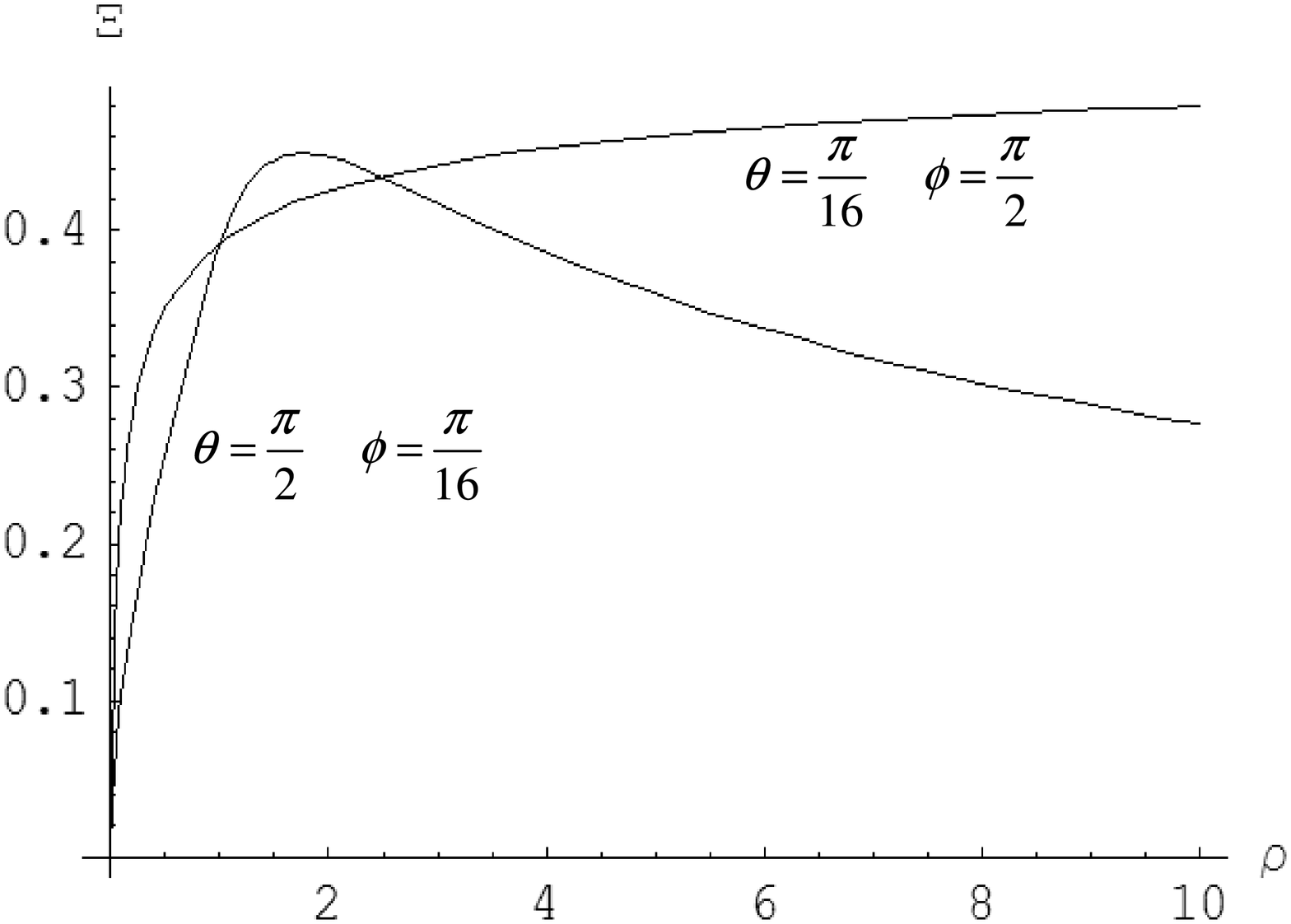}
\caption{\baselineskip = 1.0em  Lepton pair asymmetry from a pion
target.} \label{Fig.5}
\end{figure}

With the above case for the pion as a warm-up, we now proceed to lepton pair
production from a 3-quark target. For calculational purposes, it is useful
to introduce a 4-vector for the scattered hadron\cite{Vega},
\begin{eqnarray}
\xi ^{\mu } &=&(0,\cos \psi ,-i,-\sin \psi ), \\
P\cdot \xi &=&0,
\end{eqnarray}%
in terms of which the proton spinor matrix can be expressed as,
\begin{equation}
u_{\uparrow }(P)\overline{u}_{\uparrow }(P^{\,\prime })=\frac{\gamma
_{5}\,\xi \!\!/\,P\!\!\!\!/\,\,P^{^{\,\prime }}\!\!\!\!\!\!/}{2\sqrt{2P\cdot
P^{\,\prime }}}.
\end{equation}%
The form-factor contribution to lepton photoproduction off a proton is
identical to that from a pion in the limit where the spin-flip term (Pauli
form factor) is set to zero. Indeed for massless quarks and no transverse
momentum, this is strictly true. The minimal state of the proton is,%
\begin{equation}
|P_{\uparrow }\rangle =\int \frac{\left[ dx\right] }{2\sqrt{24x_{1}x_{2}x_{3}%
}}\Phi (x_{1},x_{2},x_{3})\frac{\varepsilon ^{abc}}{\sqrt{6}}u_{a\uparrow
}^{\dagger }(x_{1})\left\{ u_{b\downarrow }^{\dagger }(x_{2})d_{c\uparrow
}^{\dagger }(x_{3})-d_{b\downarrow }^{\dagger }(x_{2})u_{c\uparrow
}^{\dagger }(x_{3})\right\} |0\rangle ,
\end{equation}%
where $\left[ dx\right] =dx_{1}dx_{2}dx_{3}\delta (1-x_{1}-x_{2}-x_{3}).$
Asymptotically, $\Phi (x_{1},x_{2},x_{3})\sim 120\;x_{1}x_{2}x_{3}$ but more
realistic wavefunctions have been constructed and can be found in refs.\cite%
{Farrar},\cite{Braun}.

The proton case requires more work but it follows the calculation detailed
above for the pion. We imagine that 3 collinear quarks with charges $%
e_{1},e_{2},e_{3}$ enter from the left with momentum fractions $%
x_{1},x_{2},x_{3}$ and emerge to the right with fractions $%
y_{1},y_{2},y_{3}. $\ An extra gluon is required to transfer the hard
momentum on to the remaining quark, and this brings an additional factor of $%
g^{2}$ into the amplitude. The incoming proton is taken to be in a definite
(positive) helicity state. In the absence of quark transverse momentum, as
well as quark mass, the helicity of the final proton is that of the incoming
one. Equivalently, in this approximation, the Pauli form factor is zero.
Summing over the twelve different ways of connecting two photons to the 3
quark lines gives, after a tedious calculation:

\begin{figure}[t]
\includegraphics[height=2.0in,width=5.5in]{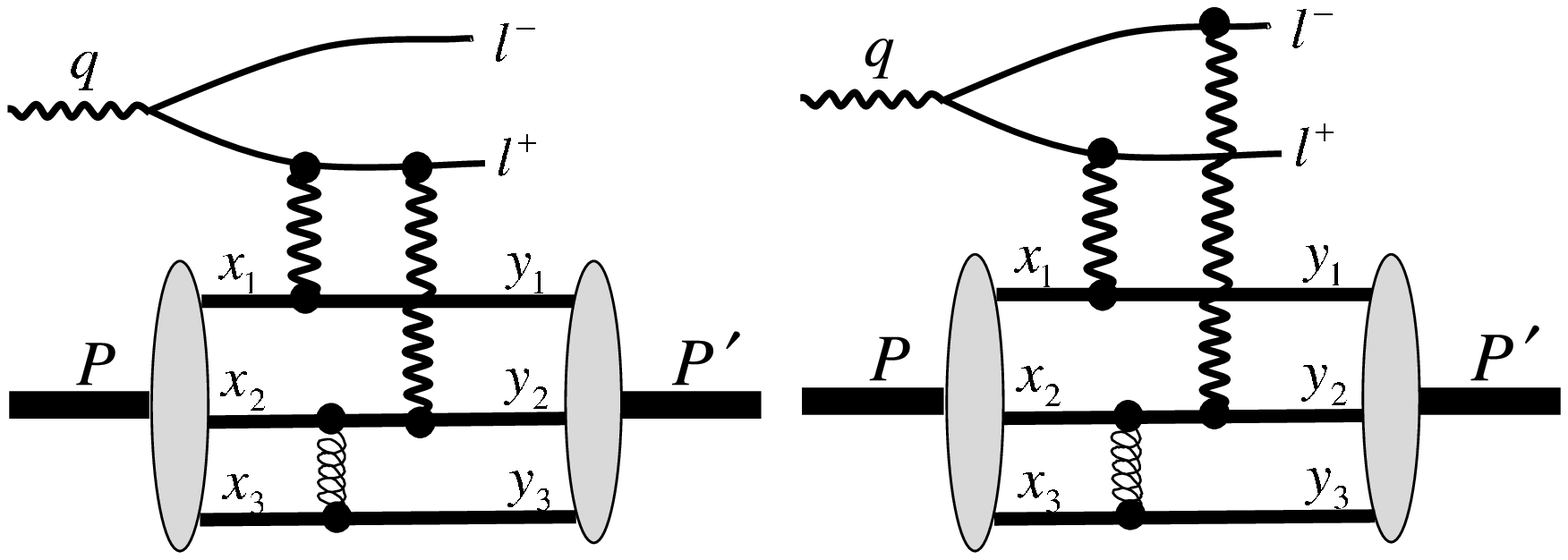}
\caption{\baselineskip = 1.0em  Typical diagrams for lepton pair
production from a 3-quark proton.} \label{Fig.6}
\end{figure}

\begin{eqnarray}
A_{2}^{\uparrow \uparrow } &\equiv &A_{2}^{\gamma N\rightarrow
Nl^{-}l^{+}}(\theta ,\phi )  \notag  \label{A2} \\
&=&\frac{2i\sqrt{2}\pi g^{2}e^{3}s\text{ }}{9M^{7}\rho ^{5/2}(1+\rho )^{2}}%
e^{-i\theta }(\cos \theta +e^{-i\phi }\frac{M}{\sqrt{-t}}\sin \theta )\left[
i\sqrt{\rho }(e^{i\theta }-1)+e^{i\phi }(e^{i\theta }+1)\right] ^{2}\times
\notag \\
&&\;\int \left[ dx\right] \left[ dy\right] \frac{%
e_{1}(e_{2}x_{2}y_{2}+e_{3}x_{3}y_{3})\Phi (x_{1},x_{2},x_{3})\Phi
(y_{1},y_{2},y_{3})x_{1}\delta (y_{1}-x_{1})}{x_{1}x_{2}x_{3}y_{1}y_{2}y_{3}%
\left[ x_{1}\overline{x}_{1}+\Lambda \right] \overline{x}_{1}^{2}}.  \notag
\\
&&
\end{eqnarray}

\begin{figure}
\includegraphics[angle=90,height=3.5in,width=5.0in]{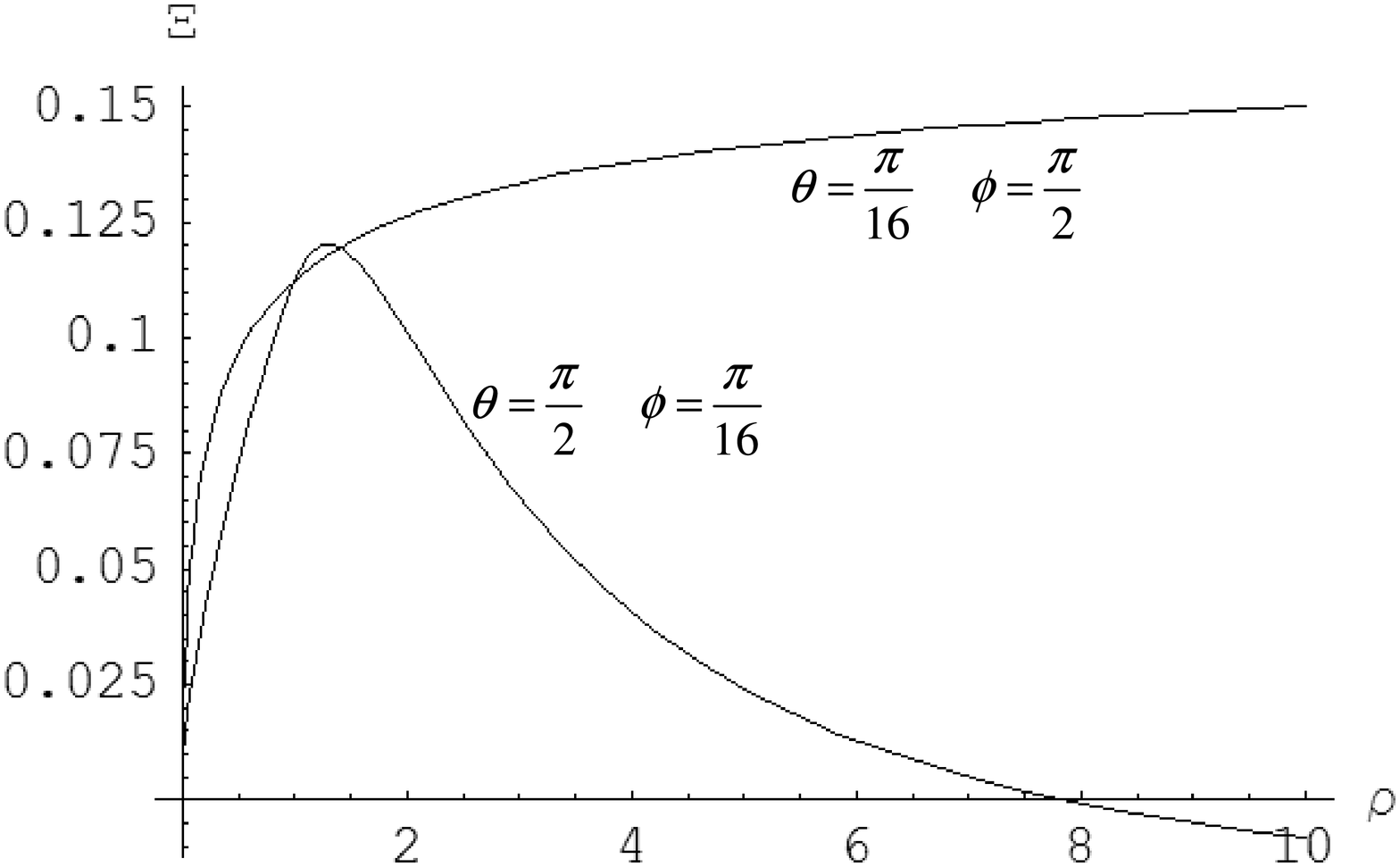}
\caption{\baselineskip = 1.0em  Lepton pair asymmetry from a proton
target.}
\label{Fig.7}
\end{figure}

In the above (as well as in the equation below), it is implicitly
understood that the complex conjugate, and exchange of labels
$(1,3)\leftrightarrows (3,1)$ are to be
added on. The charges $e_{i}$ are in units of the electron charge $e$ $%
(\alpha _{e}=e^{2}/4\pi )$. Inserting this and the single-photon results
into Eq.\ref{asy} yields,

\begin{eqnarray}
\Theta _{N}(\theta ,\phi ) &=&-\frac{64\pi ^{3}\alpha _{s}\alpha _{e}}{%
9(-t)^{2}G_{M}(t)}(\cos \theta +e^{-i\phi }\frac{M}{\sqrt{-t}}\sin \theta
)\times  \notag \\
&&\int \left[ dx\right] \left[ dy\right] \frac{%
e_{1}(e_{2}x_{2}y_{2}+e_{3}x_{3}y_{3})\Phi (x_{1},x_{2},x_{3})\Phi
(y_{1},y_{2},y_{3})x_{1}\delta (y_{1}-x_{1})}{x_{1}x_{2}x_{3}y_{1}y_{2}y_{3}%
\left[ y_{1}\overline{y_{1}}+\Lambda \right] \overline{y_{1}}^{2}}.  \notag
\\
&&
\end{eqnarray}%
As a check on the correctness of the calculations for scattering amplitudes
for both the meson and baryon cases, we have verified that the Ward identity
is satisfied in all different ways. Perturbatively $(-t)^{2}G_{M}(t)%
\longrightarrow $ constant at large $t$, and so again one has
approximate scale invariance. In Fig. 7 we have plotted the
asymmetry off a proton target. In this exploratory calculation we
have used the Chernyak-Zhitnitsky wavefunction in ref.[9]. In this
paper we have calculated asymmetries, not cross-sections because the
latter involve a 3-body phase space. Since J-Psi photoproduction has
been measured in exclusive reactions\cite{ZEUS}, lepton pair
production should also be possible.

Finally we remark that Sudakov effects, which arise from the bremstrahlung
of widely separated quarks that undergo large changes in momentum, will lead
to a weakening of the effective coupling. Thus, although the angular
structure of the amplitude will probably be similar, one must investigate
diagrams that are of one order higher in $\alpha _{s}$. For the proton case
this will involve a very large number of diagrams that will require a
machine computation. We have not attempted this calculation.\bigskip

{\Large Acknowledgments\medskip }

The author thanks Stanley J. Brodsky and Xiangdong Ji for valuable comments
and encouragement.

\end{document}